\documentclass[reprint,amsmath,amssymb,aps,]{revtex4-2}
\usepackage{bm}
\usepackage{dcolumn}
\usepackage{lineno}
\usepackage{verbatim}
\usepackage{graphicx}
\usepackage{subfigure}
\usepackage{epstopdf}
\usepackage{float}
\usepackage{threeparttable}
\usepackage{amsmath}
\usepackage{breqn}
\usepackage[colorlinks, linkcolor=blue, anchorcolor=blue, citecolor=blue, urlcolor=black]{hyperref}
\usepackage{cmap} 
\usepackage{mathtext}
\usepackage{mathrsfs}
\usepackage{makecell}

\newcommand{\pip}{\pi^{+}}
\newcommand{\pim}{\pi^{-}}

\def\piz{\pi^0}
\def\pip{\pi^+}
\def\pim{\pi^-}

\def\kstar0{K^*(892)^0}

\newcommand{\pipi}{\pi^{+}\pi^{-}}
\newcommand{\pipipi}{\pi^{+}\pi^{-}\pi^{0}}

\newcommand{\ppb}{p\bar{p}}

\newcommand{\etap}{\eta^{\prime}}

\newcommand{\pp}{\pi^+\pi^-}
\newcommand{\kk}{K^+K^-}

\newcommand{\GG}{\gamma\gamma}

\newcommand{\bfg}{\begin{figure}}
\newcommand{\efg}{\end{figure}}
\newcommand{\bitm}{\begin{itemize}}
\newcommand{\eitm}{\end{itemize}}
\newcommand{\bnum}{\begin{enumerate}}
\newcommand{\enum}{\end{enumerate}}
\newcommand{\btbl}{\begin{table}}
\newcommand{\etbl}{\end{table}}
\newcommand{\btbu}{\begin{tabular}}
\newcommand{\etbu}{\end{tabular}}
\newcommand{\bcl}{\begin{center}}
\newcommand{\ecl}{\end{center}}

\newcommand{\beq}{\begin{equation}}
\newcommand{\eeq}{\end{equation}}
\newcommand{\beqr}{\begin{eqnarray}}
\newcommand{\eeqr}{\end{eqnarray}}

\begin{document}
\title{Search for $X(1870)$ via the decay $J/\psi\to \omega\kk\eta$}
%% Saved at => 2024-04-10
\author{%%
    \begin{small}
    \begin{center}
M.~Ablikim$^{1}$, M.~N.~Achasov$^{4,c}$, P.~Adlarson$^{76}$, O.~Afedulidis$^{3}$, X.~C.~Ai$^{81}$, R.~Aliberti$^{35}$, A.~Amoroso$^{75A,75C}$, Q.~An$^{72,58,a}$, Y.~Bai$^{57}$, O.~Bakina$^{36}$
% ...
\\
\vspace{0.2cm}
(BESIII Collaboration)\\
\vspace{0.2cm} {\it
$^{1}$ Institute of High Energy Physics, Beijing 100049, People's Republic of China\\
% ...
$^{3}$ Bochum  Ruhr-University, D-44780 Bochum, Germany\\
$^{4}$ Budker Institute of Nuclear Physics SB RAS (BINP), Novosibirsk 630090, Russia\\
% ...
$^{35}$ Johannes Gutenberg University of Mainz, Johann-Joachim-Becher-Weg 45, D-55099 Mainz, Germany\\
$^{36}$ Joint Institute for Nuclear Research, 141980 Dubna, Moscow region, Russia\\
% ...
$^{57}$ Southeast University, Nanjing 211100, People's Republic of China\\
$^{58}$ State Key Laboratory of Particle Detection and Electronics, Beijing 100049, Hefei 230026, People's Republic of China\\
% ...
$^{72}$ University of Science and Technology of China, Hefei 230026, People's Republic of China\\
% ...
$^{75}$ University of Turin and INFN, (A)University of Turin, I-10125, Turin, Italy; (B)University of Eastern Piedmont, I-15121, Alessandria, Italy; (C)INFN, I-10125, Turin, Italy\\
$^{76}$ Uppsala University, Box 516, SE-75120 Uppsala, Sweden\\
% ...
$^{81}$ Zhengzhou University, Zhengzhou 450001, People's Republic of China\\
\vspace{0.2cm}
$^{a}$ Deceased\\
% ...
$^{c}$ Also at the Novosibirsk State University, Novosibirsk, 630090, Russia\\
% ...
}\end{center}
\vspace{0.4cm}
\end{small}
}
%% ends here %%

\date{\today}

%\linenumbers

\begin{abstract}
Using a sample of $(10087\pm 44)\times10^{6}$ $J/\psi$ events collected by the BESIII detector at the BEPCII collider, we search for the decay  $X(1870)\to\kk\eta$ via the $J/\psi\to \omega\kk\eta$ process for the first time. No significant $X(1870)$ signal is observed. The upper limit on the branching fraction of the decay $ J/\psi\to \omega X(1870) \to\omega\kk\eta$  is determined to be  $9.55\times 10^{-7}$ at the $90\%$ confidence level. In addition, the branching faction  $B(J/\psi\to\omega\kk\eta)$ is measured to be $(3.33\pm0.02(\rm{stat.})\pm 0.12(\rm{syst.}))\times 10^{-4}$.
\end{abstract}
\maketitle

\section{Introduction}\label{sec:introduction}\vspace{-0.3cm}
Within the framework of the Standard Model, the strong interaction is described by Quantum Chromodynamics, which predicts the existence of unconventional hadrons, such as glueballs, hybrid states and multiquark states. The discovery and characterization of such states remain a primary focus in hadron physics. The decays of the $J/\psi$ provide an excellent platform for investigating light hadron spectroscopy and searching for unconventional hadrons. Several resonances in the mass range of 1.8 to 1.9 GeV/$c^2$ have been observed in the $J/\psi$ decays, including the $X(\ppb)$~\cite{Xpp1,Xpp2,Xpp3}, $X(1835)$~\cite{X1835-1,X1835-2,X1835-3, X1835-4}, $X(1810)$~\cite{X1810-2} and $X(1870)$~\cite{X1870}.
The $X(1835)$ is first observed in
the BESII experiment in the $\pi^+\pi^-\eta^{\prime}$ invariant mass spectrum through the radiative decay of $J/\psi\to\gamma\pi^+\pi^-\eta^{\prime}$ with a statistical significance of 7.7$\sigma$~\cite{X1835-1}. This observation is  confirmed by BESIII with a statistical significance that is larger than 20$\sigma$~\cite{X1835-2}. The spin-parity of the $X(1835)$ is determined to be $0^{-+}$ in the decay of $J/\psi\to\gamma K^0_S K^0_S\eta$~\cite{X1835-3}.  In Ref.~\cite{X1835-4}, a search for the  $J/\psi\to\omega X(1835)\to\omega \pi^+\pi^-\eta^{\prime}$ hadronic decays is performed and no significant signal is observed. The $X(1870)$ resonance is first observed in the $\pipi\eta$ invariant mass spectrum via the decay of $J/\psi\to\omega\pipi\eta$~\cite{X1870} with a statistical significance of 7.2$\sigma$, based on a sample of $(225.2\pm 2.8)\times10^{6}$ $J/\psi$ events collected by the BESIII experiment. A high-statistics data sample collected with BESIII provides an opportunity to confirm the existence of the $X(1870)$ and can be searched for in other decay modes. Searching for the $X(1870)$ in the $K^+K^-\eta$ decay mode via $J/\psi\to \omega\kk\eta$ is of interest, which can provide more information on the strange quark component of the $X(1870)$.

\section{BESIII Detector}
The BESIII detector~\cite{Detector,energy1,energy2} records symmetric $e^+e^-$ collisions 
provided by the Beijing Electron Positron Collider II (BEPCII) storage ring~\cite{ring}
in the center-of-mass energy range from 2.0 to 4.95~GeV,
with a peak luminosity of $1.1 \times 10^{33}\;\text{cm}^{-2}\text{s}^{-1}$ 
achieved at $\sqrt{s} = 3.773\;\text{GeV}$.  The cylindrical core of the BESIII detector covers 93\% of the full solid angle and consists of a helium-based
 multilayer drift chamber~(MDC), a plastic scintillator time-of-flight
system~(TOF), and a CsI(Tl) electromagnetic calorimeter~(EMC),
which are all enclosed in a superconducting solenoidal magnet
providing a 1.0~T(0.9 T in 2012) magnetic field.
The solenoid is supported by an
octagonal flux-return yoke with resistive plate counter muon
identification modules interleaved with steel. 
The charged-particle momentum resolution at $1~{\rm GeV}/c$ is
$0.5\%$, and the 
$dE/dx$
resolution is $6\%$ for electrons
from Bhabha scattering. The EMC measures photon energies with a
resolution of $2.5\%$ ($5\%$) at $1$~GeV in the barrel (end cap)
region. The time resolution in the TOF barrel region is 68~ps, while
that in the end cap region is 110~ps.  
The end cap TOF system was upgraded in 2015 using multigap resistive plate chamber
technology, providing a time resolution of 60~ps, which benefits 83\% of the data used in this analysis~\cite{60ps1,60ps2,60ps3}.

\section{Dataset and MC Simulation}\label{sec_dataset}
The results reported in this article are based on a sample of $(10087\pm 44)\times10^{6}$ $J/\psi$ events~\cite{eventJ/psi} collected by the BESIII detector. 

Monte Carlo (MC) simulated data samples produced with a {\sc geant4}-base~\cite{geant4,boost} software package, which
includes the geometric description of the BESIII detector and the
detector response~\cite{energy2,response2,response3}, are used to determine detection efficiencies
and to estimate backgrounds.
To thoroughly investigate potential backgrounds, we utilize an inclusive Monte Carlo (MC) sample comprising 10 billion $J/\psi$ events. The inclusive MC sample includes both the production of the $J/\psi$
resonance and the continuum processes incorporated in {\sc kkmc}~\cite{kkmc1,kkmc2}. All particle decays are modeled with \textsc{evtgen} \cite{evtgen} using branching fractions~(BFs) either taken from the Particle Data Group (PDG)~\cite{PDG}, when available, or otherwise estimated with \textsc{lundcharm} \cite{LUNDCHARM}. Final state radiation~(FSR)
from charged final state particles is incorporated using the {\sc photos} package~\cite{photos2}.

In this study, two exclusive MC samples are employed to determine detection efficiencies. These samples correspond to the decays $J/\psi\to\omega\kk\eta$ and $J/\psi\to \omega X(1870) \to\omega\kk\eta$, each consisting of $1\times10^{7}$ MC events. The decay $\omega\to\pipipi$ is simulated using a generator considering its Dalitz plot distribution~\cite{generator}, while other decays are generated with the phase space model.

\section{Measurement of BF of $J/\psi\to\omega\kk\eta$}
\begin{figure*}[htbp]
\begin{center}
\subfigure{\includegraphics[width=\columnwidth]{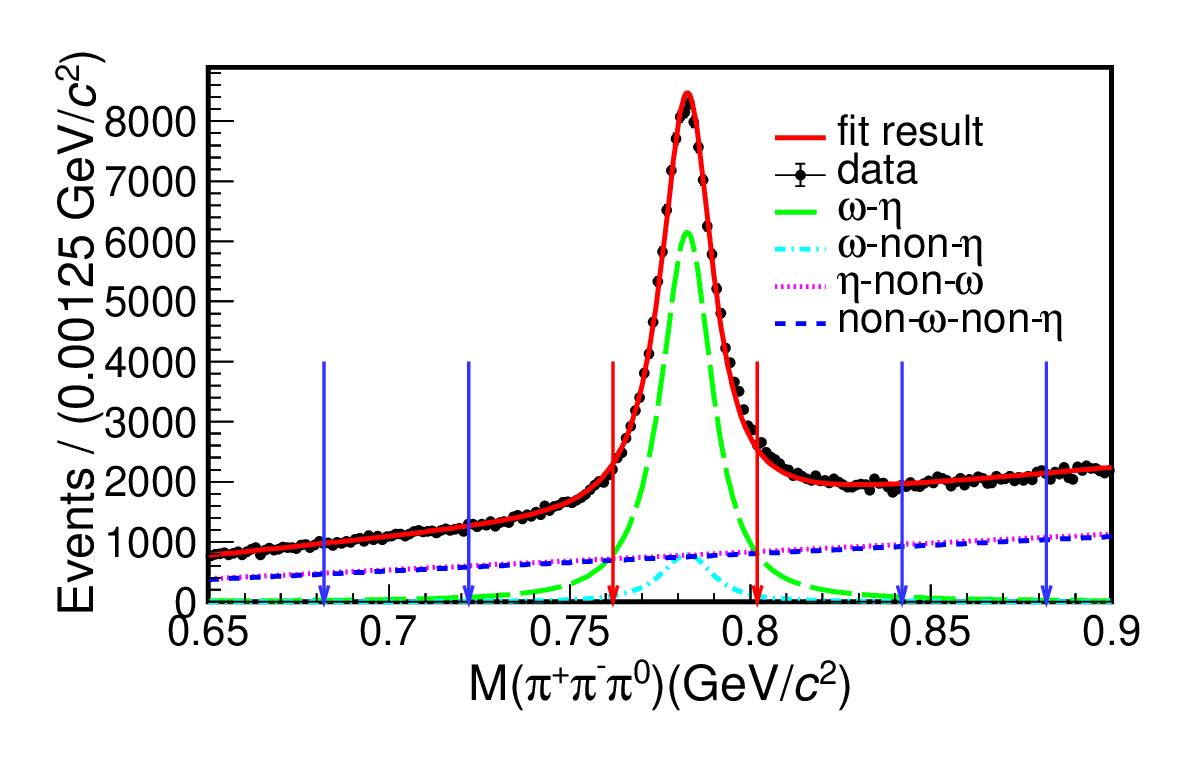}\label{fig_fitC0}}
\put(-195,130){(a)}
\subfigure{\includegraphics[width=\columnwidth]{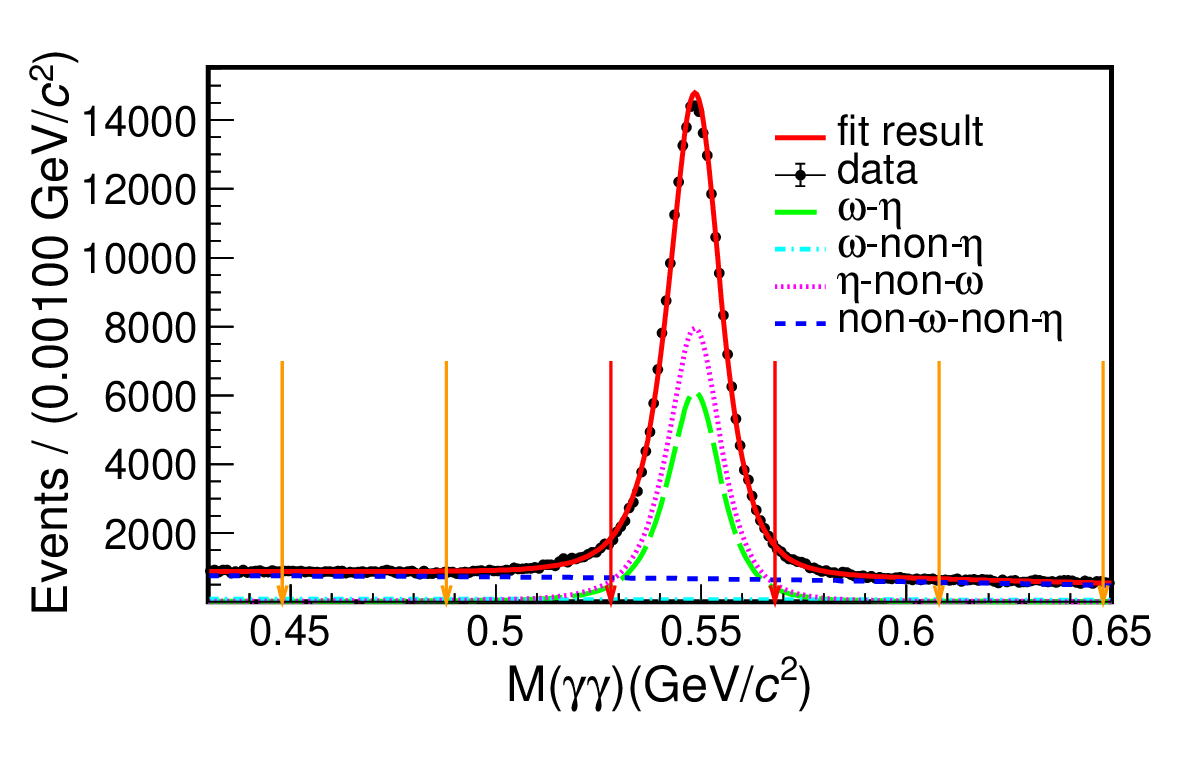}\label{fig_fitC1}}
\put(-195,130){(b)}\\
\vspace{-15pt} 
\caption{\small Projections of the 2D fit on (a)~$M(\pi^+\pi^-\pi^0)$ and (b)~$M(\gamma\gamma)$ of the accepted candidates for $J/\psi \to K^+K^-\pi^+\pi^-\pi^0\gamma\gamma$. 
 The dots with error bars are data and the red solid lines represent the fit result. The green dash lines denote the  $\omega-\eta$  signal shape, the cyan lines represent  the $\omega-{\rm non-}\eta$ peaking backgrounds, the dash pink lines denote the $\eta{-\rm non-}\omega$  peaking backgrounds and the blue lines are the ${\rm non-}\omega-{\rm non-}\eta$ peaking backgrounds. The blue arrows and yellow arrows point to the sideband regions of $\omega$ and $\eta$, respectively. The red arrows point to the signal regions of $\omega/\eta$.}
\label{fig:2dfit}
\end{center}
\end{figure*}

\subsection{Event Selection and Background Analysis}
	
The decay $J/\psi\to \omega\kk\eta$ is reconstructed with $\omega\to\pipipi$ and $\eta(\piz)\to\GG$.  The final state consists of $\kk\pp\GG\GG$, requiring four charged tracks with a net zero charge. 
Charged tracks detected in the MDC are required to be within a polar angle ($\theta$) range of $|\cos \theta|<0.93$, where $\theta$ is defined with respect to the $z$-axis,
which is the symmetry axis of the MDC. For each track, the distance of closest approach to the interaction point (IP) 
must be less than 10\,cm along the $z$-axis, $|V_{z}|$,  and less than 1\,cm in the transverse plane, $|V_{xy}|$.
Particle identification~(PID) for charged tracks combines measurements of the energy deposited in the MDC~(d$E$/d$x$) and the flight time in the TOF to form likelihoods $\mathcal{L}(h)~(h=p,K,\pi)$ for each hadron $h$ hypothesis.
Charged tracks with $\mathcal{L}(K)>\mathcal{L}(p)$ and $\mathcal{L}(K)>\mathcal{L}(\pi)$ are
identified as kaons, and those with
$\mathcal{L}(\pi)>\mathcal{L}(K)$ and $\mathcal{L}(\pi)>\mathcal{L}(p)$ as pions.

Photon candidates are identified using isolated showers in the EMC.  The deposited energy of each shower must be more than 25~MeV in the barrel region ($|\cos \theta|< 0.80$) and more than 50~MeV in the end cap region ($0.86 <|\cos \theta|< 0.92$). 
To exclude showers that originate from charged tracks, the angle subtended by the EMC shower and the position of the closest charged track at the EMC
must be greater than 10 degrees as measured from the IP. 
To suppress electronic noise and showers unrelated to the event,  the difference between the EMC time and the event start time is required to be within [0, 700] ns. The number of photon candidates in an event is at least four.

To improve the mass resolution, a kinematic fit is applied under the hypothesis $J/\psi\to K^+K^-\pi^+\pi^-\gamma\gamma \gamma\gamma$, imposing constraints on four-momentum conservation and requiring the invariant mass of one pair of photons to the nominal $\pi^0$ mass, which is called five-contraint (5C) fit~\cite{Yan:2010zze}.
 The 5C kinematic fit loops over all $\kk\pipi\GG\GG$ combinations and the one with the minimum $\chi^2_{\rm 5C}$ value is retained. 
To further suppress background contributions and improve the significance of signal, the $\chi^2_{\rm 5C}$ requirement is optimized using a figure of merit~\cite{fom} defined as $S/\sqrt{S+B}$, where $S$ denotes the number of signal events from MC simulation, and $B$ represents the number of background events estimated with $\omega$ and $\eta$ sidebands in the data. 
The nominal criterion is set as  $\chi^2_{\rm 5C} < 80$.

The mass windows for $\omega$ and $\eta$ are set to $3\sigma$ around their respective nominal masses, corresponding to 
$|M(\pipipi) - m(\omega)| < 0.02$ GeV/$c^2$ and $|M(\gamma\gamma) - m(\eta)| < 0.02$ GeV/$c^2$, where $m(\omega)$ and $m(\eta)$ are obtained from the PDG~\cite{PDG}. 
\label{sec:inclusive_selections}
To suppress backgrounds containing $\etap$, an additional requirement of $|M(\pip\pim\eta)-m(\etap)|>0.025$  GeV/$c^2$ is applied. 

To investigate 
potential background contributions, the same selection criteria are applied to an inclusive MC sample of $10\times10^9 J/\psi$ events. The topology analysis of the inclusive MC sample is performed with the generic tool TopoAna~\cite{topo}. A detailed study indicates that there is no peaking background with both an $\omega$ and an $\eta$. Three main types of background contributions are identified: the first type of background is due to $ J/\psi \rightarrow \omega K^{-} K^{*+} ,
\omega \rightarrow \pi^{0} \pi^{+} \pi^{-} , K^{*+} \rightarrow \pi^{0} K^{+} $, with an $\omega$ no $\eta$.  The second type of background is due to $ J/\psi \rightarrow \pi^{-} \eta \bar{K}^{*} K^{*+} , \eta \rightarrow \gamma \gamma , \bar{K}^{*} \rightarrow \pi^{+} K^{-} , K^{*+} \rightarrow \pi^{0} K^{+} $, with an $\eta$ but no $\omega$. The third type of background is due to $J/\psi \rightarrow \rho^{-} \bar{K}^{*} K^{*+} ,\rho^{-} \rightarrow \pi^{0} \pi^{-} ,\bar{K}^{*} \rightarrow \pi^{+} K^{-} ,K^{*+} \rightarrow \pi^{0} K^{+} $, without any  $\omega$ or any $\eta$. These contributions would be considered in the fitting for signal extraction.

\subsection{Measurement of Branching Fraction}
\label{BF}
A two-dimensional~(2D) maximum likelihood fit is performed on the distribution of $M(\pi^+\pi^-\pi^0)$ versus $M(\gamma\gamma)$ of the accepted candidates for $J/\psi \to K^+K^-\pi^+\pi^-\pi^0\gamma\gamma$ to obtain the signal yield of $J/\psi \to \omega K^+K^-\eta$. The fitting model is constructed as 

    \begin{equation}
\begin{aligned}
F= & N_{\mathrm{sig}} \times\left(F_{\mathrm{sig}}^\omega \cdot F_{\mathrm{sig}}^{\eta}\right) \\
& +N_{\mathrm{bkg}}^{\mathrm{non}-\eta } \times\left(F_{\mathrm{sig}}^\omega \cdot F_{\mathrm{bkg}}^{\mathrm{non}-\eta }\right) \\
& +N_{\mathrm{bkg}}^{\mathrm{non}-\omega} \times\left(F_{\mathrm{bkg}}^{\mathrm{non}-\omega} \cdot F_{\mathrm{sig}}^{\eta }\right) \\
& +N_{\mathrm{bkg}}^{\mathrm{non}-\omega \eta } \times\left(F_{\mathrm{bkg}}^{\mathrm{non}-\omega} \cdot F_{\mathrm{bkg}}^{\mathrm{non}-\eta }\right).
\end{aligned}
    \end{equation}

\noindent The signal shapes for $\omega$ (i.e. $F^\omega_{\rm sig}$)  and $\eta$ (i.e.  $F^\eta_{\rm sig}$) are modeled with a sum of two Johnson functions~\cite{Johnson} sharing the same mean and width parameters. The mean and width parameters for the $\omega$  and $\eta$ are 
determined from the 2D fits to the data. The tail parameters and fractions of each  Johnson function are fixed to the values obtained from the fit to the signal MC events. The background shapes of non-$\omega$ (i.e. $F^{\mathrm{non}-\omega}_{\rm bkg}$) and non-$\eta$ (i.e. $F^{\mathrm{non}-\eta}_{\rm bkg}$) are described by first-order and second-order Chebychev polynomial functions with free parameters, respectively. $N_{\rm sig}$ is the number of signal events, while $N_{\mathrm{bkg}}^{\mathrm{non}-\eta } $, $N_{\mathrm{bkg}}^{\mathrm{non}-\omega}$ and $N_{\mathrm{bkg}}^{\mathrm{non}-\omega \eta }$ represent the event numbers of the three types of backgrounds mentioned above.  The projections of the 2D fit on $M(\pi^+\pi^-\pi^0)$ and $M(\gamma\gamma)$ are shown in Fig.~\ref{fig:2dfit}.
 The BF of $J/\psi\to\omega\kk\eta$  is calculated by
\begin{equation}
B(J/\psi\to \omega\kk\eta) = \frac{N_{\rm sig}}{N_{J/\psi} \cdot B_{\rm int} \cdot \epsilon},
\label{eq:brul}
\end{equation}

\noindent where $\epsilon=9.98\%$ is the detection efficiency obtained by MC simulation, $N_{J/\psi}$ is the number of $J/\psi$ events in the data sample, and $ B_{\rm int}$ is the product of the BFs for
 $\omega \rightarrow \pipipi$, $\pi^{0} \rightarrow \gamma\gamma$ and $\eta \rightarrow \gamma\gamma$ quoted from the PDG~\cite{PDG}. 
The signal yield from the 2D fit is $N_{\rm sig}=116136\pm504$.  The BF is determined to be $B(J/\psi\to\omega\kk\eta)=(3.33\pm0.02(\rm{stat.}))\times10^{-4}$.

 \subsection{\label{sec:exclusive_selections} Systematic Uncertainties}

The systematic uncertainties on the BF measurement are from tracking, PID, photon detection,  number of $J/\psi$ events, quoted BFs, 5C kinematic fit, background rejection, 2D fit and MC simulation. Details are discussed below. 

The systematic uncertainties associated with $\pi^\pm$  tracking and PID are evaluated using a control sample of $J/\psi \rightarrow p\bar{p}\pi^{+}\pi^{-}$. The efficiency differences between data and MC simulation for the control sample are used to reweight the signal MC sample. The systematic uncertainties tracking and PID of two pions are both taken as $1.7\%$.  Similarly, the systematic uncertainties of tracking and PID of two kaons ($K^\pm$) are assigned as $0.5\%$ and 0.1\%, respectively, using a control sample of $e^+e^- \rightarrow \pi^+\pi^- J/\psi$ with $J/\psi \rightarrow K^+K^-K^+K^-$.

The systematic uncertainty related to the photon detection is studied using a control sample of $e^+e^-\to\gamma\mu^+\mu^-$~\cite{photocorrect1}.  
The relative difference of 1.4\% in the momentum reweighted efficiency between data and MC simulation is assigned as the systematic uncertainty.

The systematic uncertainty from the number of $J/\psi$ events is $0.4\%$ according to Ref.~\cite{eventJ/psi}. The quoted BFs~\cite{PDG} of   $\omega \rightarrow \pipipi$, $\pi^{0} \rightarrow \gamma\gamma$ and $\eta \rightarrow \gamma\gamma$ are $(89.2\pm0.7)\%$, $(98.823\pm0.034) \%$ and $(39.36\pm0.18) \%$, respectively.  The quadratic sum of the individual contributions, 0.9\%,  is assigned as the total systematic uncertainty due to the quoted BFs.

The systematic uncertainty from the kinematic fit  is estimated by correcting the helix parameters of the charged tracks in the MC simulation~\cite{Helix}. The differences in the detection efficiencies with and without the corrections for the helix parameters, 0.8\% is the taken as the uncertainty. 

The systematic uncertainty from the $\eta^\prime$ veto is estimated by varying its veto range within $\pm 1\sigma$. The maximum change of 0.9\% in BF is assigned as the systematic uncertainty.

To estimate the systematic uncertainties related to the signal shapes, an approach based on MC simulations has been used. Two thousand sets of MC simulation samples, with an equivalent size as data, are generated based on the nominal fitting results. Each of them is fitted with both the nominal and alternative signal shapes. The alternative signal shape is modeled with a sum of two Johnson functions~\cite{Johnson} and a Crystal Ball function~\cite{CB}. The relative differences in the signal yields between the nominal fit and alternative are calculated. This distribution of the differences is then fitted with a Gaussian distribution. The mean values of individual Gaussian distributions, 0.5\% and 0.3\%, are taken as the systematic uncertainties for the $\omega$ and $\eta$ signal shapes, respectively.

To evaluate the systematic uncertainties from the background shapes, the same method as that for the signal shape is applied. The background shapes are changed from the first-order to a second-order Chebychev polynomial function for the non$-\omega$ backgrounds, and from the second-order to a third-order Chebychev polynomial function for the non$-\eta$ backgrounds. The systematic uncertainties due to the background shapes for non$-\omega$ and non$-\eta$ are determined to be 0.3\% and 0.4\%, respectively.

To take into account the difference between data and MC simulation in the invariant mass distributions of $K^+K^-$, the invariant mass spectra are divided into 10 bins, and an averaged efficiency is calculated from the signal MC, by weighting the efficiency obtained for each bin by the fraction of generated events for each bin; in analogous way the averaged efficiency is calculated for the data sample, by weighting the selected events using the  bin-dependent efficiency obtained from MC. The difference between the averaged efficiency for simulation and for data is taken as the systematic uncertainty of MC model.  

The systematic uncertainties are summarized in Table~\ref{tab:sysbr}. Each source of systematic uncertainty is treated as individual value and summed in quadrature.
	
 \begin{table}[!htbp]
\centering
\caption{Relative  systematic uncertainties on the BF measurement for $J/\psi\to\omega\kk\eta$.}
\label{tab:sysbr}
 \setlength{\tabcolsep}{4mm}
 \resizebox{\linewidth}{!}{
\begin{tabular}{lcc}
\hline
\hline
Source	& Uncertainty (\%)  \\
\hline
$\pi^\pm$ tracking	        & 1.7     \\
$K^\pm$ tracking 	            & 0.5      \\
 $\pi^\pm$ PID     & 1.7      \\
$K^\pm$ PID	                    & 0.1    \\
Photon selection 			    & 1.4        \\
Number of $J/\psi$ events  	    & 0.4       \\
Quoted BFs 				& 0.9       \\
Kinematic fit			& 0.8      \\
Veto of $\etap$                 & 0.9     \\
Signal shape                   &   0.5     \\
Background shape	      &  0.4      \\
MC model                       &1.5  \\
 \hline
Total				      &  3.6  \\
\hline
\hline
 \end{tabular}
 }
 \end{table}

\section{\label{sec:exclusive_selections} Search for $X(1870)$ in $J/\psi\to \omega X(1870) \to\omega\kk\eta$}
Furthermore, we search for the $X(1870)$ resonance in the distribution of the $K^+K^-\eta$ invariant mass, $M(K^+K^-\eta)$, based on the selected candidates for  $J/\psi \to \omega K^+K^-\eta$.
	
\subsection{Background Analysis}
\begin{figure}
\hspace*{0.55cm}
\centering
\subfigure{\includegraphics[width=0.45\textwidth]{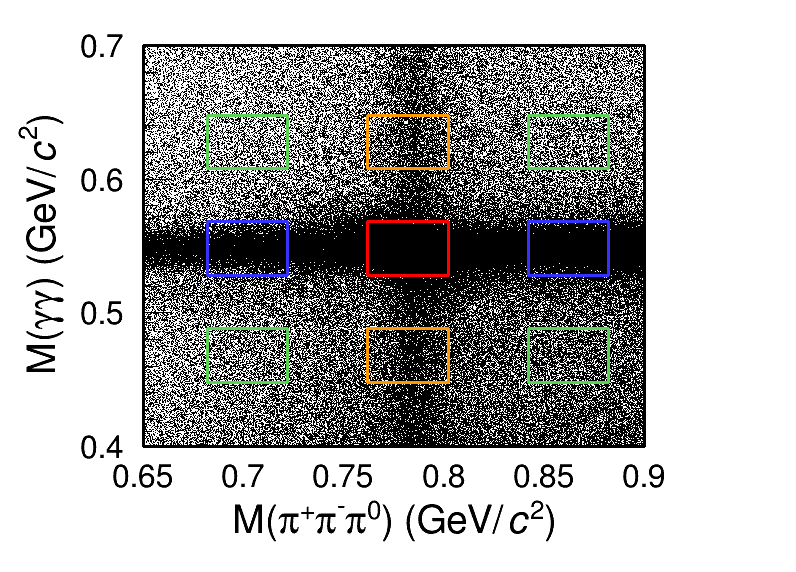}}

\vspace{-15pt} 

\caption{\small The distribution of $M(\pipipi)$ versus $M(\gamma\gamma)$ of the accepted candidates for $J/\psi \to K^+K^-\pi^+\pi^-\pi^0\gamma\gamma$ in data. }
\label{fig:2dsideband}
\end{figure}
 		
Detailed topology analysis with the inclusive MC sample within the mass range of $[1.7, 2.1]$ GeV/$c^2$ of the $\kk\eta$ mass spectrum indicates that there is no peaking background with both $\omega$ and $\eta$ in the final states. To estimate the background contribution, we use a data-driven approach that utilizes 2D sideband regions of $\omega$ and $\eta$. The sideband regions of $\omega/\eta$ are defined as 0.06~GeV/$c^2$$<|M(\pipipi)-m(\omega)|<0.1~$GeV/$c^2$ and  0.06 GeV/$c^2$$<|M(\gamma\gamma)-m(\eta)|<0.1$ GeV/$c^2$, corresponding to $(7-13)\sigma$ away from the $\omega$ or $\eta$ nominal masses. 
Figure~\ref{fig:2dsideband} shows the  distribution of $M(\pipipi)$ versus $M(\gamma\gamma)$. The regions A is indicated with green solid line boxes, while the regions B is marked with yellow solid line boxes and the regions C with blue solid line boxes. The number of background events in the signal region, denoted as $N_{\rm bkg}$, is estimated as $N_{\rm bkg}=0.50N_{\rm B} +0.53N_{\rm C} -0.265N_{\rm A}$, where $N_{\rm A}$, $N_{\rm B}$ and $N_{\rm C}$ represent the number of events in the regions A, B, and C, respectively. The normalization factors for events in the sideband regions are estimated by the 2D fit on
$M(\pi^+\pi^-\pi^0)$ and $M(\gamma\gamma)$ of the accepted candidates for $J/\psi \to K^+K^-\pi^+\pi^-\pi^0\gamma\gamma$ in data. The fitting model for the 2D fit and the fit results are as described above. The background fraction is estimated to be $28.5\%$.

\subsection{Upper Limit of the Branching Fraction for $X(1870)$}
\begin{figure}[htbp]
    \hspace*{-0.8cm} 
    \centering
    \includegraphics[width=0.45\textwidth]{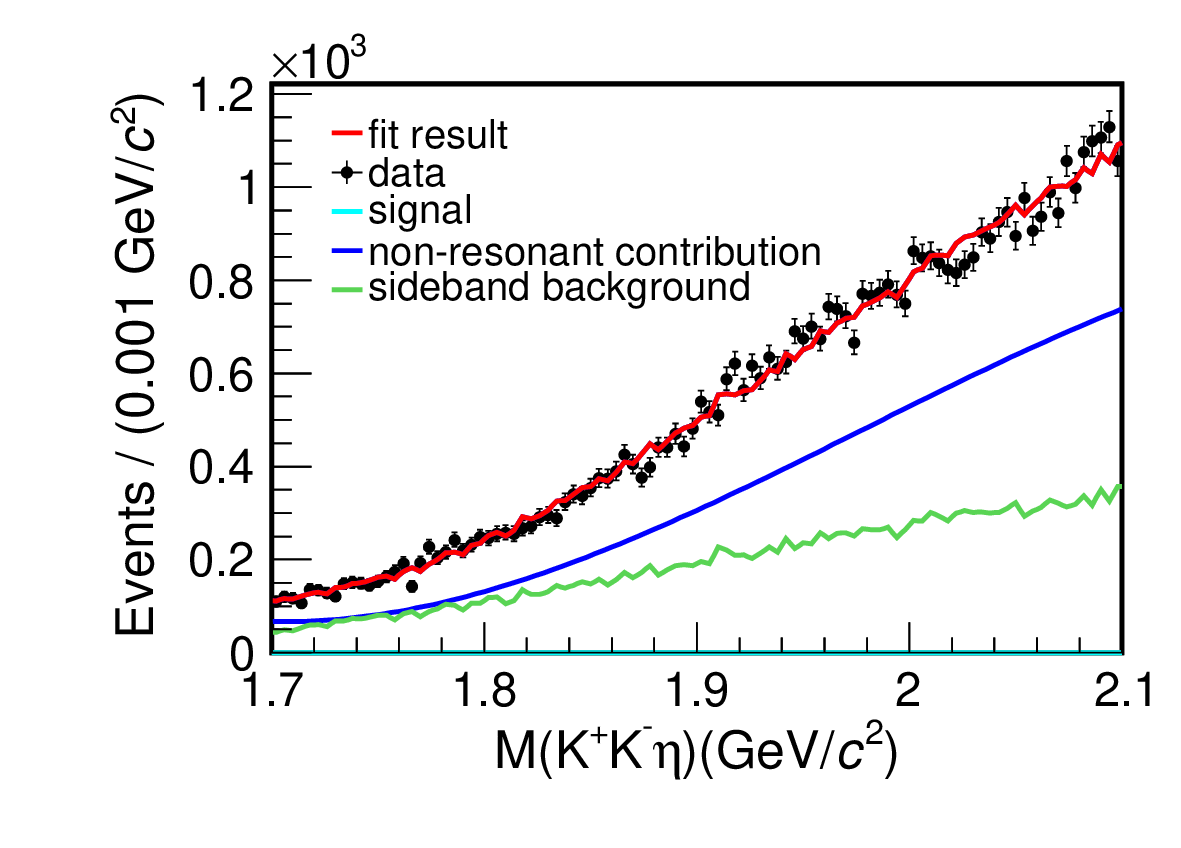}
    
    \vspace{-16pt} 
    
    \caption{\small The fit to the $M(K^+K^-\eta)$ distribution of the accepted candidates for $J/\psi \to \omega K^+K^-\eta$. The dots with error bars are data, the cyan line represents the fitted signal shape, the green line denotes the 2D $\omega/\eta$ sideband background from data and the blue line represents other non-peaking backgrounds.}
    \label{fit.fit}
\end{figure}

To search for $X(1870)$ via the decay $J/\psi\to \omega X(1870) \to\omega\kk\eta$, the maximum likelihood fit is performed to the $M(K^+K^-\eta)$ distribution of the accepted candidates for $J/\psi \to \omega K^+K^-\eta$. In the fit, it is assumed that there is no interference between the $X(1870)$ and non-$X(1870)$ components. The signal shape is described by a Breit-Wigner function defined as
 \begin{equation}
 {\rm BW}(s)=\frac{1}{M_{\rm R}^2-s-i M_{\rm R} \Gamma_{\rm R}},
 \end{equation}
where $M_{\rm R}$ and $\Gamma_{\rm R}$ are the mass and width of the $X(1870)$. The width of the Breit-Wigner function is fixed to  0.057  GeV/$c^2$, and the mass is fixed to 1.8773 GeV/$c^2$~\cite{X1870}. $\sqrt{s}$ is the $K^+K^-\eta$ invariant mass. The background contributions are estimated with the $\omega/\eta$ 2D sidebands. The non-resonant contribution is described by a free third-order Chebychev polynomial function. The fit result is shown in Fig.~\ref{fit.fit}, where the cyan line represents the fitted $X(1870)$ signal. Since no $X(1870)$ signal is observed, the upper limit on the number of $X(1870)$ signal events is determined at the 90\% confidence level~(C.L.). The details are described in the next section.

\subsection{Systematic Uncertainties}
\label{sec:sysU}

In the search for $J/\psi\to \omega X(1870) \to\omega\kk\eta$, the
systematic uncertainties are categorized into additive and multiplicative 
uncertainties. 
The additive uncertainties originate from the fit to extract the signal yield. The uncertainty in the signal shape is studied by changing the Breit-Wigner function to the MC simulated shape. The systematic uncertainty related to the width and mass of $X(1870)$ is  estimated by varying the nominal mass and width by  $\pm1\sigma$~\cite{X1870}. To account for the systematic uncertainty arising from the 2D sideband backgrounds, the number of events in the 2D sideband backgrounds is varied within $\pm1\sigma$ and  the sideband shape is varied by shifting the sideband ranges within $\pm1\sigma$. The systematic uncertainty from the non-peaking background  is examined with an alternative fit with a second-order Chebychev polynomial function.  The
resulting upper limits for each case is determined, and the maximum value is taken
as the upper limit. 

The multiplicative uncertainties are associated with the efficiencies, and will affect the BF calculation.
The systematic uncertainties from the tracking and PID, the photon selection, the number of $J/\psi$ events and the quoted BFs are the same as those mentioned above.

The systematic uncertainty associated with the $\omega$ or $\eta$ signal region selection is estimated by varing the $\omega$ and $\eta$ signal regions. The relative differences in efficiencies between data and MC simulation, $1.2\%$ for $\omega$ and $1.6\%$ for $\eta$, are taken as the systematic uncertainties.

The systematic uncertainty in the quantum numbers of the $X(1870)$ is evaluated by assuming it as a pseudoscalar meson. The resulting 7.7\% change in efficiency is considered as the systematic uncertainty.

A difference of 1.4\% in efficiency with and without correcting the helix parameters in the 5C kinematic fit is taken as the systematic uncertainty due to the kinematic fit.

The systematic uncertainty associated with the $\etap$ veto is
studied by varying the $\etap$ veto range within $\pm 1\sigma$ of its fitted resolution.  The maximum difference of 1.0\% in the BF is take as the uncertainty.

The multiplicative uncertainties on the BF measurement for $J/\psi\to \omega X(1870) \to\omega\kk\eta$ are  summarized in Table~\ref{tab:sysup}. The total multiplicative systematic uncertainty is obtained by summing the individual contributions in quadrature.

\begin{table}[!htbp]
\centering
\caption{The multiplicative uncertainties for the upper limit on the BF measurement for $J/\psi\to \omega X(1870) \to\omega\kk\eta$.}
\label{tab:sysup}
\setlength{\tabcolsep}{4mm}
\begin{tabular}{>{\fontsize{10}{12}\selectfont}lcc} 
\hline
\hline
Source						     & 	Uncertainty (\%)\\
\hline
$\pi^\pm$ tracking		            &  1.7\\
$K^\pm$ tracking                    &  0.5\\
$\pi^\pm$ PID	                     &  1.7\\
$K^\pm$ PID                         &  0.1\\
Photon selection 			         &  1.4  \\
Number of $J/\psi$ events     & 0.4\\
Quoted BFs 				           &  0.9 \\
Kinematic fit			             &   1.4\\
Veto of $\etap$                     & 1.0\\
$\omega$ signal region               & 1.2     \\
$\eta$ signal region                 & 1.6     \\
Quantum number of $X(1870)$         & 7.7  \\
\hline
Total						           &  8.7 \\
\hline
\hline
\end{tabular}
\end{table}

To incorporate the multiplicative systematic uncertainties in the calculation of the upper
limit, the likelihood distribution is convolved by a Gaussian function with a mean of zero and a width equal to $\sigma_{\epsilon}$ as described in Refs.~\cite{Bayesian,ul1,ul3,ul2} with 

\begin{dmath}
    L'(B) \propto \int_{0}^{1} L(B\frac{\epsilon}{\epsilon_0})e^{\frac{-(\epsilon-\epsilon_0)^2}{2\sigma^2_\epsilon}}d\epsilon,
    \label{eq:Ln}
\end{dmath}

\noindent where $L(B)$ is the likelihood distribution as a function of the yield $n$, $\epsilon_{0}$ is the detection efficiency and $\sigma_{\epsilon}$ is the multiplicative systematic uncertainty. The upper limit on the BF at the 90\% C. L., $B^{\rm U.L.}_{\rm sig}$, is obtained by integrating the likelihood function to 90\% of its physical region.  
Finally,
with the detection efficiency~($\epsilon^{\prime}$) of $ 7.02\%$  obtained from MC simulation, the upper limit on the BF of the signal decay at the 90\% C.L. is set to be $9.55\times 10^{-7}$. 

\section{Summary}\label{sec:summary}
Based on the sample of $(10087\pm 44)\times10^{6}$ $J/\psi$ events collected from the BESIII detector,    the BF of the decay $J/\psi\to\omega\kk\eta$ is measured to be $(3.33\pm0.02(\rm{stat.})\pm 0.12(\rm{syst.}))\times10^{-4}$ for the first time. No significant $J/\psi\to\omega X(1870)\to\kk\eta$ signal is observed.  The upper limit on the product BF of the decay $ J/\psi\to \omega X(1870) \to\omega\kk\eta$  at the $90\%$ C. L. is determined to be $9.55\times 10^{-7}$ for the first time.
 In Ref.~\cite{X1870}, the $X(1870)$ resonance has a clear signal in the $\pipi\eta$ invariant mass spectrum. However, there is no evidence of $X(1870)$ in the $\kk\eta$ invariant mass spectrum.  It suggests that the $X(1870)$ may have a limited $s$-quark content. 
%These results provide important information to further understand the nature of the $X(1870)$, which can be further investigated by studying additional $J/\psi$ decay modes.
To understand the nature of $X(1870)$, it is
critical to measure its spin and parity and to search for
it in more decay modes with higher statistics of $J/\psi$ data
samples in the future.

%% Saved at => 2024-04-10
\textbf{Acknowledgement}

The BESIII Collaboration thanks the staff of BEPCII and the IHEP computing center for their strong support. This work is supported in part by National Key R\&D Program of China under Contracts Nos. 2020YFA0406300, 2020YFA0406400, 2023YFA1606000; National Natural Science Foundation of China (NSFC) under Contracts Nos. 11635010, 11735014, 11935015, 11935016, 11935018, 12025502, 12035009, 12035013, 12061131003, 12192260, 12192261, 12192262, 12192263, 12192264, 12192265, 12221005, 12225509, 12235017, 12342502, 12361141819; the Chinese Academy of Sciences (CAS) Large-Scale Scientific Facility Program; the CAS Center for Excellence in Particle Physics (CCEPP); Joint Large-Scale Scientific Facility Funds of the NSFC and CAS under Contract No. U1832207; 100 Talents Program of CAS; The Institute of Nuclear and Particle Physics (INPAC) and Shanghai Key Laboratory for Particle Physics and Cosmology; German Research Foundation DFG under Contracts Nos. 455635585, FOR5327, GRK 2149; Istituto Nazionale di Fisica Nucleare, Italy; Ministry of Development of Turkey under Contract No. DPT2006K-120470; National Research Foundation of Korea under Contract No. NRF-2022R1A2C1092335; National Science and Technology fund of Mongolia; National Science Research and Innovation Fund (NSRF) via the Program Management Unit for Human Resources \& Institutional Development, Research and Innovation of Thailand under Contract No. B16F640076; Polish National Science Centre under Contract No. 2019/35/O/ST2/02907; The Swedish Research Council; U. S. Department of Energy under Contract No. DE-FG02-05ER41374.

%% ends here %%

\end{document}